\documentclass[%
 reprint,
 amsmath,amssymb,
 aps,
]{revtex4-2}

\usepackage{graphicx}% Include figure files
\usepackage[utf8]{inputenc}
\usepackage{float}
\usepackage{dcolumn}% Align table columns on decimal point
\usepackage{bm}% bold math
\DeclareUnicodeCharacter{00A0}{ }
\begin{document}

\preprint{}

\title{Spin-Orbit Proximity Effect in Bi/Co Multilayer: The Role of Interface Scattering}% Force line breaks with \\

\author{Arthur Casa Nova Nonnig}
 \affiliation{Instituto de Física, Universidade Federal do Rio Grande do Sul, 90501-970 Porto Alegre, RS, Brazil}%Lines break automatically or can be forced with \\
\author{Alexandre da Cas Viegas}%
\affiliation{Instituto de Física, Universidade Federal do Rio Grande do Sul, 90501-970 Porto Alegre, RS, Brazil}%
\author{Fabiano Mesquita da Rosa}
\affiliation{Instituto de Física, Universidade Federal do Rio Grande do Sul, 90501-970 Porto Alegre, RS, Brazil}%
\author{Paulo Pureur}
\affiliation{Instituto de Física, Universidade Federal do Rio Grande do Sul, 90501-970 Porto Alegre, RS, Brazil}%
\author{Milton Andre Tumelero}
 \email{matumelero@if.ufrgs.br}
\affiliation{Instituto de Física, Universidade Federal do Rio Grande do Sul, 90501-970 Porto Alegre, RS, Brazil}%

\date{\today}% It is always \today, today,
             %  but any date may be explicitly specified

\begin{abstract}
The Spin-Orbit Proximity Effect is the raise of Spin-Orbit Coupling at a layer near to the interface with a strong spin-orbit material. It has been seen in several system such as graphene and ferromagnetic layers. The control of the Spin-Orbit Coupling can be a pathway to discover novel and exotic phases in superconductor and semimetallic systems. Here, we study the magnetoelectrical transport, i.e., magnetoresistance and anomalous Hall effect, in Cobalt/Bismuth multilayers looking for traces of spin-orbit proximity effect and evaluate the origin of such effect. Our results point for an increase of Spontaneous Magnetic Anisotropy of Resistivity and Anomalous Hall Resistivity at very low thicknesses of Cobalt. The analysis of the Anomalous Hall Resisitivity indicate that the Bismuth layers change the scattering mechanism of Hall effect to the extrinsic skew-scattering type, indicating that the spin-orbit proximity effect could be related to the elastic scattering of cobalt free carriers by bismuth sites at the interface. 

\begin{description}

\item[Keywords]
Spin-Orbit Proximity Effect, Ferromagnetism, Bismuth Multilayers.
\end{description}
\end{abstract}

%\keywords{Suggested keywords}%Use showkeys class option if keyword
                              %display desired
\maketitle

%\tableofcontents

\section{Introduction}

The interplay between Spin-Orbit Coupling (SOC) and Ferromagnetism (FM) has been shown to be responsible by several novel and emergent phenomena in physics, most of them discovered in the last decades \cite{Manchon2015,Rev2017}. Such discoveries, such as Skyrmions \cite{Fert2013,fert2017}, Quantum Anomalous Hall Effect \cite{aqha1,aqhe3}, Spin-Orbit Torque \cite{SOCtorque1,SOCtorque2,SOCtorque3} and Triplet Magnetic Josephson Junctions \cite{Linder2019,SOCtripletSC1} became even more outstanding
because of its promising applications in information technology \cite{Manchon2015}. Most of this emergent physics occurs, or is enhanced, at the interfaces between FM and strong Spin-Orbit Materials (SOM) \cite{Rev2017}. From the technological perspective, one of the main objectives is to increase or control the SOC in semiconductor and ferromagnetic systems in order to improve the materials applicability. Some current attempt are based in ferroelectric tuning of Rashba SOC \cite{SOC-FE1,SOC-FE2} and Proximity Induced SOC \cite{socproximity,socprox2}. Although the Ferroelectricity can help to control the SOC in semiconductors, it can hardly be used for metallic FM system. 

The proximity induced SOC or Spin-Orbit Proximity Effect (SOPE) occurs in few atomic layers of a low spin-orbit coupling  (SOC) metal near its interface with a material characterized by strong SOC. It was observed when interfacing Graphene with strong SOC materials \cite{socprox2,socproxPt1} or by doping Graphene with strong SOC impurities \cite{socproxPt2}. In graphene-based heterostructures, the effect essentially results from the hybridization of the $\pi$ orbital in graphene with the d-band of metallic layers at interfaces \cite{SOCprox-Au, socprox2,socproxPt2,socproxPt1,socprox1}. Hybridization of the $\pi$ orbital with localized wave-functions of point defects is the basis for SOPE induced by impurities \cite{socproximity}. Such hybridization processes remove the spin degeneracy without destroying the Dirac cones \cite{socproxPt2}. Also interesting is the fact that he Rashba SOC in graphene could also be enhanced by codoping. In this process, the presence of p- and n- dopants could lead to the presence of an electrostatic potential that breaks the inversion symmetry \cite{socprox-pn}. 

In the case of SOM/FM bilayers, the inverse effect, known as the magnetic proximity effect, has been observed, in which the SOM presents magnetic properties \cite{bi2se3-moodera,magprox}. The case of proximity enhancement of SOC in the FM layers, it was reported by Zhang, et al \cite{Zhang2015}. A 10-fold increment of SOC in CoFeAl (CFA) layers at CFA/Pt bilayer samples was observed by measuring the Anomalous Hall Effect (AHE). However, the origin the SOPE remains unclear in FM material. Zhang, et al, \cite{zhang2013} suggested the possibility of magneto-scattering of the charge carries to play as a source for the enhacement of SOC in CFA/Pt bilayer. It seems clear that the mechanism responsible for SOPE, whether governed by hybridization or scattering processes, should be first fully understood in order to allow its application in spintronics devices quantum computing, etc.

Here we study the SOPE in ultrathin Cobalt and Bismuth multilayers. We used Bi as a source of SOC at the Bi/Co interfaces and measured SOC sensitive effects such as Spontaneous Magnetic Anisotropy of the Resistivity (SAR) and Anomalous Hall Effect (AHE) in order to map the SOC at the Cobalt Layers. SAR describes the resistivity anisotropy resulting from the relative orientation of the electrical current and the magnetization. This effect is also called as anisotropic magnetoresistance (AMR) since it is measured in the presence of small fields applied either parallel or perpendicular to the current. SAR furnishes a direct measure of the SOC intensity \cite{campbellchapter}. The anomalous Hall resistivity, otherwise, scales with magnetization, thus being intrinsically dependent on the spin-orbit interaction. The Bismuth was chosen due to it strong SOC and low conductivity. Our findings shown evidences of a enhancement of SOC at the Co layers close to the interface with bismuth. Our results suggested that the origin of the observed SOC increment is related to spin-dependent scattering of conduction electrons of Co by the Bismuth atoms at the interfaces. 

\section{Methods}

The samples were prepared by Magnetron Sputtering using independent targets of Bismuth and Cobalt (both 99.99$\%$). Power sources with 10W (DC) and 45W (RF) were employed for Bi and Co sputtering, respectively. The Vacuum chamber base pressure was $ 8 x10^{-8} torr $, the deposition pressure was $ 1 x10^{-3} torr $, maintained by 200 $cm^3/mim$ (sccm) of ultrapure argon gas (99.9999 $\%$) flux. The substrate was a 100 nm $SiO_{2}$ over a silicon (100) wafer. The sample thicknesses were confirmed by Rutherford Backscattering and the crystal structure was checked by X-Ray Diffraction, showing polycrystalline pattern for both Bismuth and Cobalt (in HCP crystal structure). All the samples were covered by a 2 nm Titanium capping layer to prevent degradation. Samples with the following structure were prepared $[Co(t)Bi(3)]_{x10}$ with t = 0.5, 0.75, 1.0, 2.0 nm. The Magnetoresistance and Hall effect were measured using a standard Hall Bar geometry, with six electrical contacts. Current was applied at the extremities of the samples. Two inner contacts were used for detecting the longitudinal resistivity signal, while two other, placed in opposite edges of the sample, were used for measuring the Hall voltage. For the magnetoresistance experiments,
in-plane and perpendicular-to-plane magnetic field orientations were applied. Fields up to 1.5 kOe were used for the magnetoresistance measurements, while fields up to 40 kOe were employed for the Hall effect experiments.  Temperatures were varied in the range 3 K to 300 K. The magnetotransport experiments were performed using an equipment based on a CMag9 cryocoller manufactured by Cryomagnetics Inc.. Auxiliary magnetization measurements were also carried out in the low field range. For these experiments, a VSM EV9 magnetometer manufactured by Microsense was used.

\section{Results}

\begin{figure}
	   \includegraphics[scale=0.6]{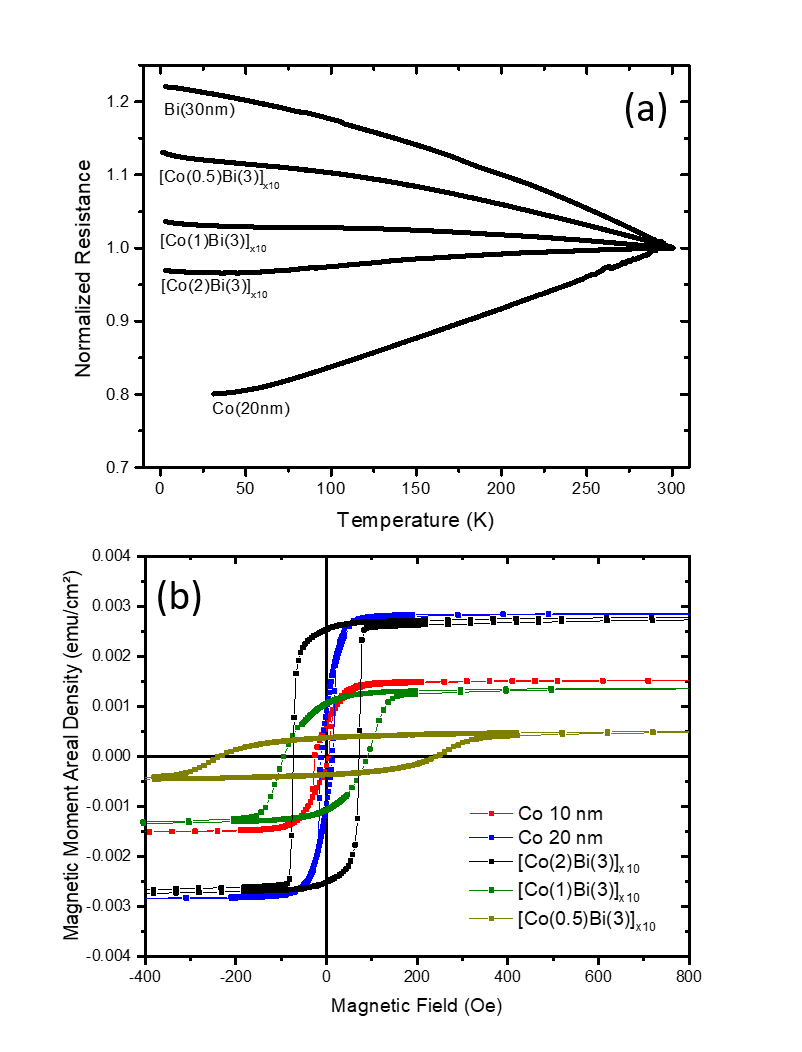} 
	   \caption {In panel (a) are shown the normalized (at room temperatura) electrical resistivity vs. temperature for the studied Co/Bi Multilayer. Also shown are measurements for the reference samples Bi(30nm) and Co(20nm). In (b) The Magnetization vs. Magnetic Field, the Hystereses curve for the multilayered samples measured at 300 K whith the field laying at the samples plane.}
	   \label{f1}
\end{figure}

\emph{Electrical and magnetic characterization}. Figure \ref{f1}a shows the normalized electrical resistance of the studied Co/Bi multilayers as a function of the temperature. Results for the references samples, Bi(30 nm) and Co(20 nm) are also presented. The Co reference sample displays the typical metallic behavior with Residual Resistance Ratio (RRR) of about 1.2, which is low compared to values reported in refs. \cite{rovan2020,aheCo1}, but expected due to very low thicknesses of the layers. The Bismuth otherwise shows a Negative Temperature Coefficient (NTC) which is related to the polycrystalline nature of the layers. It is worthwhile to mention that the absolute value of Bi(30nm) sample resistivity is two orders of magnitude higher than that for Co(20nm). The electrical behavior of the multilayers cannot be simply described by a parallel or series association of resistors representing the individual layers. In films with small Cobalt thickness, such as 0.5 and 0.75 nm, the layers are not necessarily continuous, but form large flat clusters, which imposes a serial-like electrical behavior to the heterostructure. For thicker Co layers, such as  1 and 2 nm, the transport is likely to flow mostly within the Co layers in a parallel-type association with the Bi acting as spacer. As displayed in Figure \ref{f1}a, a change in the resistivity vs. temperature behavior, from Positive Temperature Coefficient (PTC) to NTC is observed at the Bi/Co content ration gradually increases. 

The Figure \ref{f1}b shows the magnetic behavior of the samples in the M x H curve. The saturation magnetization confirm that the magnetic moment of Cobalt layers does not change significantly due to stacking with Bismuth layers. In the same figure, one also observes a systematic increase of coercivity by intercalating Cobalt with Bismuth. The coercive field $H_c$ sistematically increases with the thickness of the Bi layers. This behavior could be related to the increase of the spin-orbit coupling.

\begin{figure}
	   \includegraphics[scale=0.4]{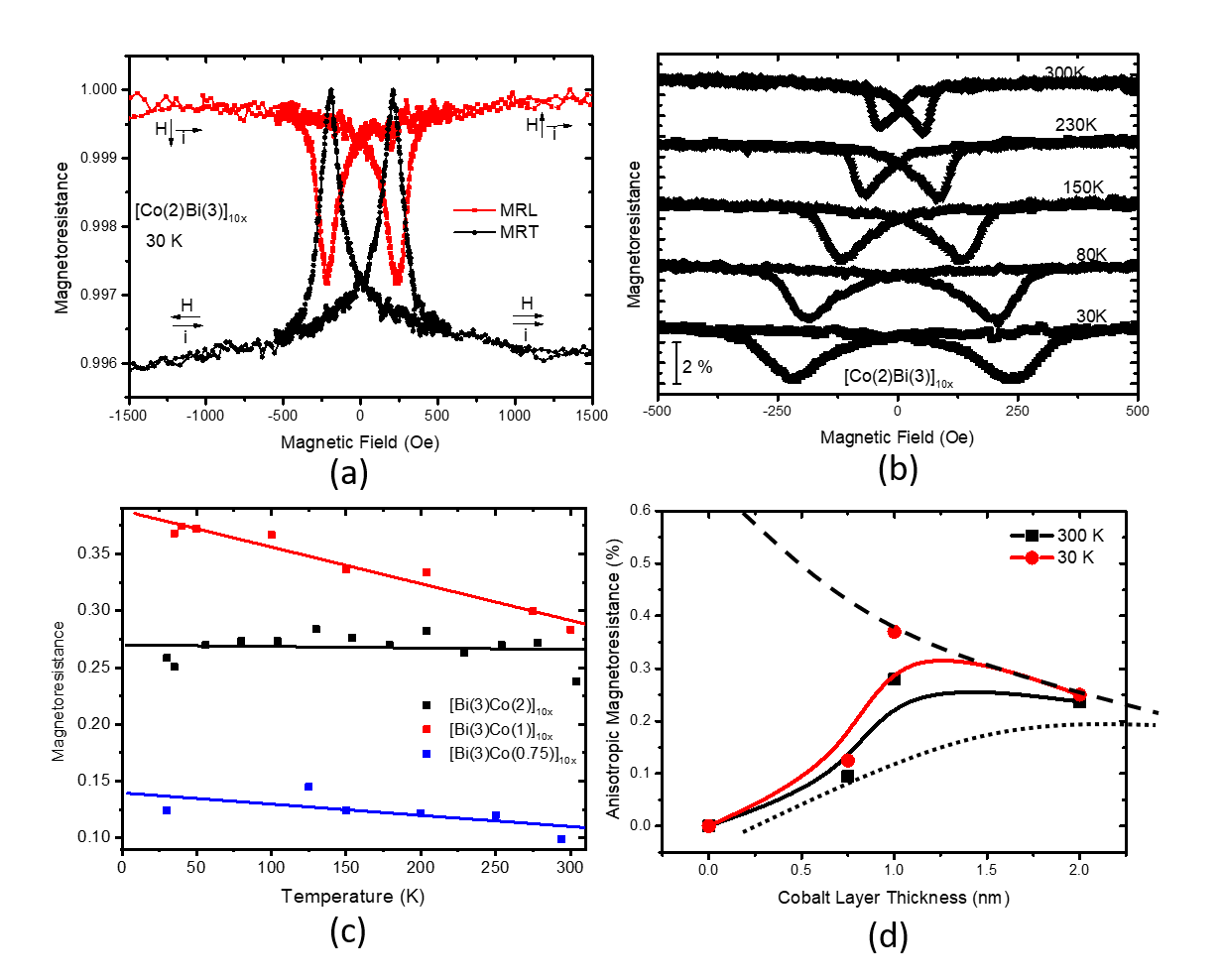} 
	   \caption {In (a) the magnetoresistance curve of the samples $[Co(2)Bi(3)]_{x10}$ in longitudinal configuration (MRL) is shown in black, while the transversal configuration (MRT) is shown in red. (b) Magnetoresistance for the samples $[Co(2)Bi(3)]_{x10}$ measured at the quoted temperatures. (c) The SAR ratio vs. temperature and (d) the SAR ratio vs. Cobalt thickness measured at 30 K and 300 K. The contínuous black and red lines in this panel are guide for the eyes. The dashed and dotted lines are explained in the text. }
	   \label{f2}
\end{figure}

\emph{Magntotransport characterization}. Now we proceed to analyse the results of magnetotransport of the Co/Bi multilayers. The magnetoresistance of the sample $[Co(2)Bi(3)]_{x10}$) is depicted in the Figure \ref{f2}a. These measurements were done at 30 K, under two distinct field configuration, namely, transversal (MRT) and longitudinal (MRL), in the first, both magnetic field and electrical current lays horizontally in the sample plane, but are perpendicular to each other. In the case of longitudinal, the field and current are parallel to each other. The opposite sign of magnetoresitance in the configuration MRT and MRL are signature of the (SAR) \cite{inverseAMR,Campbell_1970} in ferromagnetic materials. This effect comes from the Cobalt layers only. The arrows in the Figure \ref{f2}a indicate the magnetization and current relative orientations. The MRL and MRT values could depends on the samples magnetic anisotropy and do not are necessarily equals  \cite{AMRFeNi}. 

Figure \ref{f2}b shows the SAR curves for the samples $[Co(2)Bi(3)]_{x10}$ obtained at different temperatures. Notice that there is almost no change in the AMR ratios by varying the temperature. The decrease of the temperature, nevertheless, increase in the coercive field, as seen from the peak separation. The same increase in coercivity is observed in the other samples with Bismuth intercalation, but not in the pure cobalt samples. 

The SAR ratio is given by \cite{Campbell_1970}
\begin{equation}
    \Delta \rho / \rho_{(0)}= \gamma(\alpha-1)
\end{equation}
where $\rho_0$(0) is the resistivity at zero field, $\gamma$ is a parameter depending on the spin-orbit coupling and $\alpha = \rho_{\uparrow}/\rho_{\downarrow}$ is the ratio between the resistivities for the two spin-dependent current channels. $\Delta \rho$ could be experimentally determined as $\Delta \rho = MRL - MRT$, where both longitudinal and transverse magnetoresistances are obtained at the technical saturation field. Either $\gamma$ and $\alpha$ are not expected to depend much on the temperature. Thus, the results in Figure \ref{f2}b are in agreement with expectations from the Campbell-Fert \cite{Campbell_1970} model for SAR.   

The Figure \ref{f2}c presents the SAR (or AMR) Ratio vs. Temperature for the samples $[Co(t)Bi(3)]_{x10}$ with t = 0.75, 1 and 2 nm. As discussed above, the SAR ratio barely depends on the temperature, except for the sample $[Co(1)Bi(3)]_{x10}$ in which this ratio decreases weakly, but consistently, upon heating. This behavior is possibly due to a mixing of the spin-dependent currents, leading to a decrease in the effective value for alpha in Eq(1). A similar effect was also observed in magnetic Platinum layers (due to the magnetic proximity effect) \cite{magprox}. Figure \ref{f2}d shows the SAR ratio for the studied multilayers versus the cobalt thickness. The highest SAR ratio was found for the sample $[Co(1)Bi(3)]$. The peak profile shown by the red and black lines (for 30 and 300 K, respectively) may be interpreted as originated by two factors. The first one is the bismuth shunt at very low thicknesses of Co \cite{liu2010}, which fully drives the electrical current through the Bismuth layers, nullifying the SAR ratio at the lowest Co thickness. Such effect is represented by the dotted line in the Figure \ref{f2}d. The peak of SAR in \ref{f2}d suggests the occurrence of proximity effect due to the Bismuth layers, which increases locally (near to the interface) the SOC and consequently the SAR ratio. In the limit of large contents of Co in the multilayer, the relative effect of the proximity effect should become less pronounced leading to a decrease in the SAR ratio. The expected variation of SAR due to the Spin-Orbit Proximity Effect (SOPE) as a function of the Co thickness is displayed by the dashed line in Figure \ref{f2}d, for temperatures of 30 K and 300 K. Then, the combination of the shunt effect and SOPE lead to the round maximum observed in SAR data of Figure \ref{f2}d. Notice that the dotted and dashed lines are only a qualitative description of the effect pointed above, and were not obtained using any model, they are only eye-guides. Similar results were observed by Honda, et al, \cite{Honda2003}, in Bi-Co thin films alloys.

\begin{figure}
	   \includegraphics[scale=0.35]{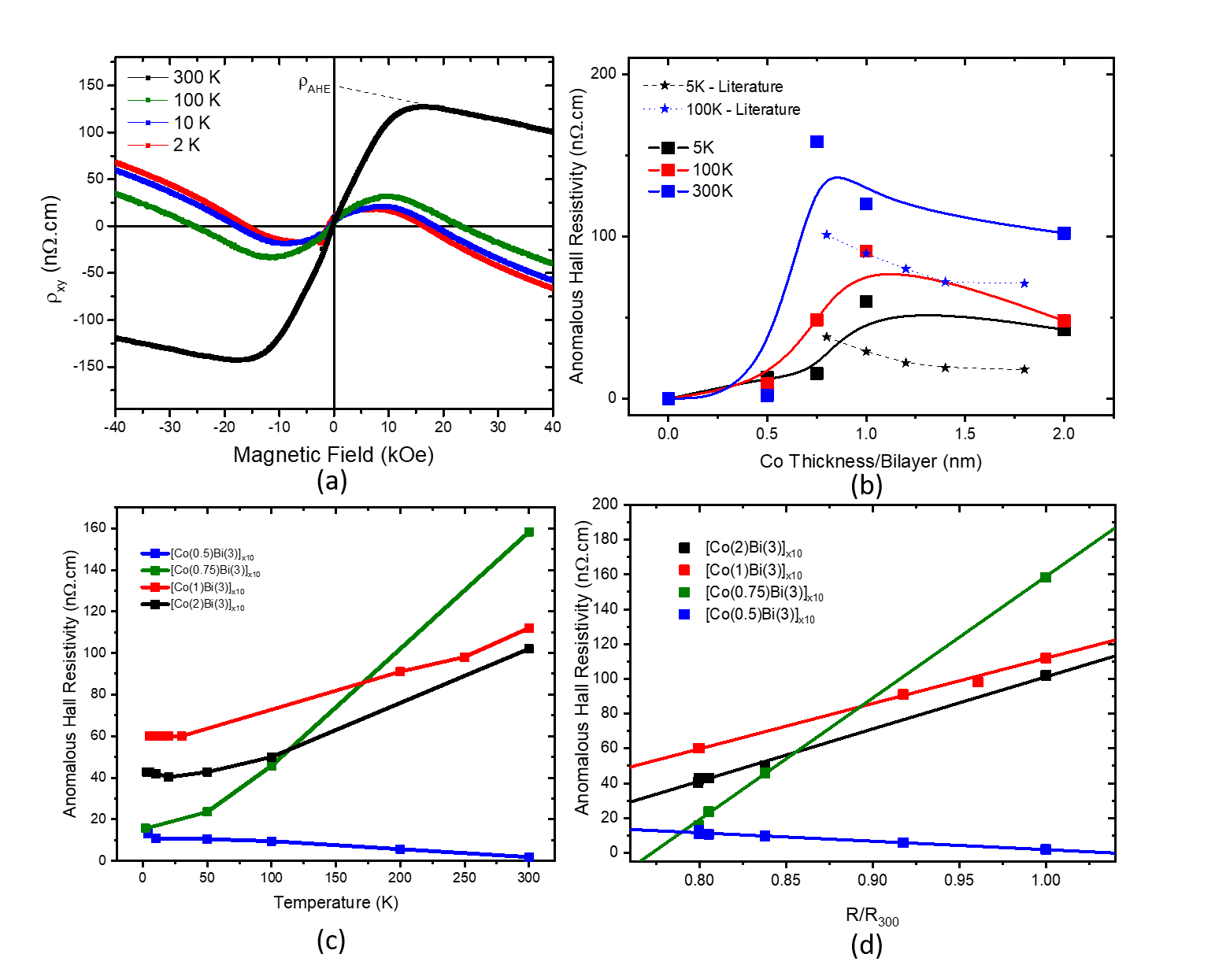} 
	   \caption {IN (a) Transverse Hall resistivity for the sample $[Co(2)Bi(3)]_{x10}$ at different temperatures vs. external magnetic field. (b) The Anomalous Hall Resistivity vs. Cobalt Thickness in the multilayer, at distinct temperatures. (c) The Anomalous Hall Resistivity vs. Temperature and (d) the Anomalous Hall Resistivity vs. the longitudinal resistivity for the multilayers.}
	   \label{f3}
\end{figure}

\emph{Anomalous Hall effect characterization}. Hall effect results for the studied samples are show in  the Figure \ref{f3}. The Figure \ref{f3}a displays the total Hall Resistivity ($\rho_{xy}$) for the sample $[Co(0.75)Bi(3)]_{x10}$ at different temperatures. The results show clearly the presence of an anomalous contribution, which scales with the magnetization, superimposed to the ordinary effect that scales with the applied field, as given by the usual expression.
\begin{equation}
    \rho_{xy} = R_{0}H + R_{S}.M
\end{equation}
where $R_{0}$ and $R_{S}$ are the ordinary and anomalous Hall coefficients, respectively. The anomalous Hall resistivity, $\rho_{AHE}$, is obtained by extrapolating the high-field linear behavior of $\rho_{xy}$ to zero field. The so determined $\rho_{AHE}$ are shown in Figure \ref{f3}b for the studied heterostructures at three different temperatures. Interestingly enough, $\rho_{AHE}$ shows the same peak profile as obtained in the SAR analysis.

 It is clear in Figure \ref{f3}b that $\rho_{AHE}$ is dependent on the temperature. This is due to the fact that $R_{S}$ varies a power law of the longitudinal resistivity for Co. The qualitative description of the data is quite the same as discussed previously for SAR. At small thicknesses there is a shunt of the Bismuth layer and the $\rho_{AHE}$ is strongly reduced. Increasing the Co content, SOPE causes a sharp increase in $\rho_{AHE}$, that goes through a maximum at about 1 nm Co, then decreases slowly with further augmentation of the Co thickness. As for SAR, the effect of the proximity induced SOC near the interface becomes less relevant in the response of the whole multilayer when the Co thickness reaches a certain threshold.  At the highest Co content, the Hall resistivity typical of bulk cobalt is recovered and a saturation is observed. This behavior is also shown by the data collected in ref\cite{aheCo1} and depicted in \ref{f3}b, in star symbols. These data are for the Hall resistivity of thin Co films with variable thickness. The observed increase of $\rho_{AHE}$ by decreasing film thickness was interpreted as effect of SOPE at the surface of Co[27]. The $\rho_{AHE}$ values found in Ref.\cite{aheCo1} are smaller than the one seen in our multilayers. This fact could be attributed to the stronger SOC induced by the Bi layers, as compared to that due to surface. Moreover, the shunt effect does not occur in the single layers of Co studied in ref ref\cite{aheCo1}. 

\begin{figure}
	   \includegraphics[scale=0.4]{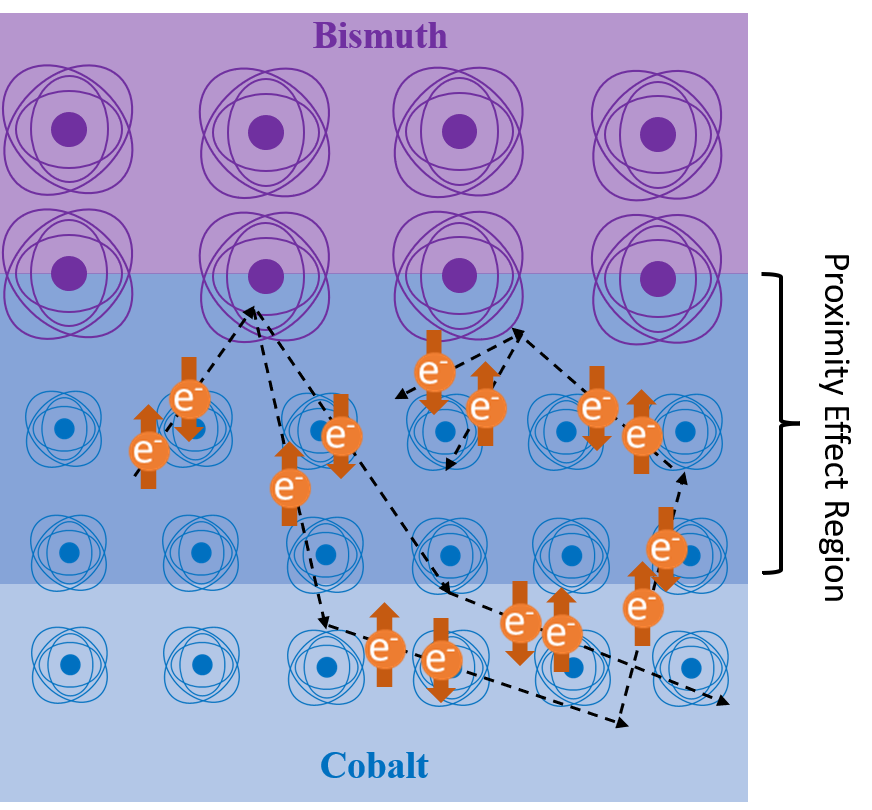} 
	   \caption {Schematic of the spin-dependent scattering near to the interface induced be the strong Bismuth spin-orbit and leading the the SOPE.}
	   \label{f4}
\end{figure}

Measurements of $\rho_{AHE}$ as a function of temperature for the studied multilayers are shown in Figure \ref{f3}c, whereas Figure \ref{f3}d depicts the $\rho_{AHE} vs. \rho_{xx}$ plot.  We point out, however, that was not possible single out the longitudinal resistivity due to the Cobalt layer of each heterostructure. Then, we used the $\rho_{xx}$ for the Co (20nm) reference sample for all the multilayers. The straight lines that relates $\rho_{AHE}$ and $\rho_{xx}$ data in Figure \ref{f3}d are consistent with a description based on the skew-scattering mechanism for the AHE observed in our multilayers \cite{Nagaosa}.  

\emph{Discussion about the AHE and SAR data}. The prevalence of this extrinsic type of scattering, i.e., the Skew Scattering, suggests that the Bi atoms at the interface are the main responsible by the enhancement of $\rho_{AHE}$ in the multilayers. Notice that the dominant AHE mechanism in bulk cobalt is the intrinsic one \cite{Luttinger1954,Nagaosa}, where a quadratic relation between $\rho_{AHE}$ and $\rho_{xx}$ is expected \cite{aheCo1,aheCo2}. Then, our data indicate that the SOPE observed in the SAR and AHE in the multilayers is related to a spin-dependent scattering of charge carriers in cobalt by the strong SOC bismuth atoms at the interfaces. Such spin-dependent scattering occurs only near to interface, given the range of the SOPE. To enlighten the discussion above, a simple scheme of the spin-dependent scattering close to the interface is presented at the Figure \ref{f4}. In this picture, the region of the Co layer affected by SOPE is indicated. 

In previous report, Zhang Y.Q., et al, \cite{Zhang2015} showed that the SOPE can be estimated as a combination of both, Skew Scatting and Side Jump contribution to Spin-Momentum scattering. Here our analysis indicate that Skew Scattering seems to be the dominant mechanism on this Bi/Co heterostructures. Sanches J.C.R., et al. \cite{Sanchez2013} showed that the introduction o Bi can lead to a high enhancement in the Inverse Spin Hall Effect signal on Ag/NiFe heterostructure. Our results can also shed some light in this debate about the spin-to-charge conversion in the Ag/Bi bilayer \cite{Sanchez2013,Ag-Bi1,Ag-Bi2}, once such system should also present SOPE due to the presence of Bi layer.

Another important discussion is the impact of the structure and morphology of the multilayers in the SAR and AHE results. Zhang Y. Q., et al. \cite{Zhang2015} pointed that the interdiffusion of atoms at the interfaces and the roughness can slight decrease the Hall angle, and so, the Anomalous Hall Resistivity. Here, we do not expect flat and uniform layers for Co thickness smaller than 1 nm, however, the presence of impurities should reduce the RRR ratio of the samples, with imply in almost no dependence of $\rho_{AHE}$ on the temperature. Which is the opposite what occurs in the results (see Figure \ref{f3}b) presented here. So we the presence of small portion of Bi interdiffusion on the Co should be of small importance to the description of our results. 

In conclusion, by performing detailed measurements of the spontaneous anisotropy of the resistivity and the anomalous Hall effect  Bi/Co multilayers we have found evidences of Spin-Orbit Proximity Effect (SOPE) occurring at Co ferromagnet thin layers near its interface with the Bi layers. Our findings indicate that the enhancement of spin-orbit coupling close to the interface between layers is related to the spin-dependent scattering of carrier in the Co layer by the strong spin-orbit coupling of Bi atoms.

\begin{acknowledgments}
We would like to acknowledge the funding agencies CAPES, CNPQ and FAPERGS by the support. We thanks the grants CNPQ/UNIVERSAL 438171/2018-7, CNPQ/PQ 312938/2020-9, CNPQ/PQ 310171/2017-2 and PRONEX 16/0490-0. We also would like to thank the support of the Laboratory of Nanoconformation (LCN) and Laboratory of Ionic Implantation (LII) at UFRGS for XRD and RBS facilities.  
\dots.
\end{acknowledgments}

\appendix

\bibliography{sope.bib}% Produces the bibliography via BibTeX.

\end{document}